\documentstyle[stwol]{article}

    [1995/07/18 v2.1 PSfrag (MCG)]
\RequirePackage{epsfig}

\def\be{\begin{equation}}
\def\ee{\end{equation}}
\def\bea{\begin{eqnarray}}
\def\eea{\end{eqnarray}}

\begin{document}

\title{THERMAL DILEPTONS AS A POSSIBLE SOURCE  OF THE
SOFT DILEPTON ENHANCEMENT MEASURED IN A-A COLLISIONS AT SPS ENERGIES}

\author{ R. BAIER, M. DIRKS }

\address{Fakult\"at f\"ur Physik, Universit\"at Bielefeld, D-33501 Bielefeld, Germany}

\author{ K. REDLICH$^a$}

\address{GSI, PF 110552, D-64220 Darmstadt, Germany and
Institute of  Theoretical Physics, University of Wroc\l aw, PL-50204 Wroc\l aw, Poland }


\twocolumn[\maketitle\abstracts{
The production of soft dileptons in a thermal mesonic medium
is discussed in the context of   recent CERN experimental data
reported by the CERES Collaboration. We do not intend to give
a general and critical review, but instead concentrate mainly
on our approach, however, incorporating many of the recent attempts in the
literature.
 We calculate  the contributions to the dilepton yield
arising from     pion annihilation
and $\pi -\rho $ scattering.
It is shown that thermal dileptons from $\pi -\rho$ scattering
give a significant         contribution to the low-mass yield, however, it
 can only partly account for the experimentally observed soft
dilepton excess  seen in S-Au and Pb-Au collisions at SPS energy.
The out off-equilibrium effects as well as a dropping vector meson mass
are  discussed in the context of the thermal dilepton yield.
 We emphasize, following the results of  Li, Ko, and
 Brown, that, until now,
 the best way
 to provide
 a quantitative explanation of the        observed enhancement of low-mass dileptons by the CERN
 experiments   is
 the assumption of a   decreasing
  vector meson mass   in a high density thermal medium.\footnote{
  Invited talk presented at the 28th International Conference on
  High Energy Physics, July 1996, Warszawa, Poland.}
 }]

\section{Introduction}
Electromagnetic radiation has been widely discussed as one of the
most sensitive signals to probe the dynamics of nuclear
 interactions.~\cite{shur,mclerran,russkanen,redl}
Recently, three  experiments  at CERN with S-beams of 200
 GeV/nucleon:  CERES~\cite{ceres,pfeiffer}, HELIOS/3~\cite{helios}
  and NA38~\cite{na38} - report
an enhanced production of dileptons  with
a yield far beyond a mere superposition of p-p or p-nucleus collisions.
Also recent measurements by the  CERES Collaboration with Pb-beams at SPS
energy show a similar behaviour of  the dilepton rate.
The excess starts at a mass $ > 200{\rm MeV/c}^2$ and persists up to the mass
of the $J/\Psi $. These results have   recently implied  very intensive theoretical discussions
in order to understand the origin of the
 excess.~\cite{dinesh,ko,cassing,vesa,wambach,our,zah,ko2,shur2,koch,kap}
The properties of the observed yield,  particularly the fact
that the enhancement starts at $M\sim  2m_\pi$
 might indicate
that the excess is just due to additional     thermal dileptons
originating from     pion annihilation.~\cite{dinesh}

The possibility of the creation of hot and dense QCD matter
in thermal equilibrium in relativistic collisions of heavy nuclei at SPS energy
has  already got             experimental as well as theoretical
support. In particular, the measured particle spectra   at
CERN experiments with heavy ion beams
 seem  to be rather well explained by a thermal model for particle production.~\cite{model}
 The appearance of a thermal hadronic medium in heavy ion collisions naturally implies
the  additional source of dileptons, absent in p-p or p-A collisions.
\begin{figure}
\center
 \psfig{figure=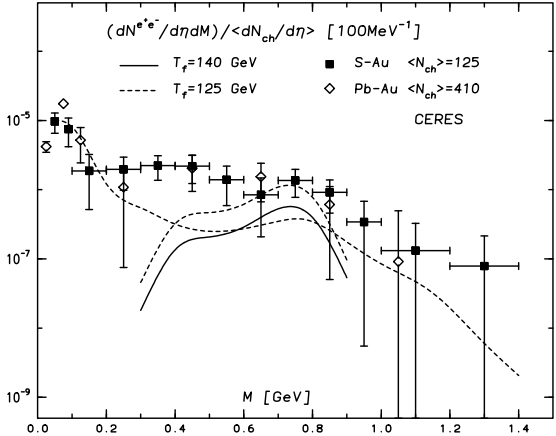,height=6.2cm}
\caption{Thermal dileptons from  $\pi^+\pi^-\to e^+e^-$
 annihilation obtained in terms of Bjorken model in comparison with CERES S-Au and Pb-Au data.
 The dashed line, calculated by CERES for S-Au, represents the expected dielectron
 yield from all known hadronic sources.}
\label{fig:1   }
\end{figure}
At     SPS energy  the thermal medium is mostly of mesonic origin
as the measured charged-pion/proton ratio is still  of  the order of five. Thus,
the most relevant process for     dilepton production
should be due to       meson  scatterings and decays. The differential
thermal rate per unit space-time volume $d^4x$
 at which mesons interacting  create $e^+e^-$ pairs
of invariant mass $M$ can be obtained either  in terms of kinetic
or thermal field theory.~\cite{mclerran,weldon}

In heavy ion collisions, the additional complication arises
as the rate per unit spacetime volume is not  an experimentally  measured  distribution.
Dileptons are produced in an expanding medium, thus to compare
the theoretical results with the experiment one still needs to
perform  the space-time integration. In our following discussion
we       apply     Bjorken$^,$s hydrodynamical model~\cite{bj} for the  expansion
dynamics.
In a mesonic medium pion annihilation with rhos and virtual photons
in the intermediate state is the basic source of thermal $e^+e^-$ pairs.
In fig.1 we compare thermal dielectron yields from pion annihilation with the CERES experimental data.
The calculation has been done in the Bjorken model~\cite{bj}
with  the initial temperature $T\sim 0.2$ GeV as arising from $dN_{ch}/dy\sim 150$
the measured yield in central S-Au collisions. The initial thermalization time is  taken as $\tau_i\sim 1$fm
and the freeze-out temperature is considered in fig.1 as a  free
parameter. The results of fig.1 confirm
 that   indeed in the vicinity of $\rho$-meson  peak
pion annihilation together with the hadronic cocktail from Dalitz
and vector meson decay  is
well compatible with the data.~\cite{dinesh,ko} This, however, requires  the
freezeout temperature in the range between 120-140 MeV, where the
mean  free path of pions is of the order of the sulphur radius.

Recently the CERES Collaboration has measured dilepton yields
in Pb-Au collisions at SPS energy.~\cite{pfeiffer} The preliminary data averaged
over all events measured with different $dN_{ch}/dy$ show
a  dilepton enhancement   by a factor $2.5\pm 0.5$ in a mass range
$0.2<M<1.5$ GeV. This is a smaller  excess  than previously
seen     for S-Au collisions.
 Thus, we have additional data
to check the validity of a thermal model at least near
   the $\rho$-meson peak. The comparison of a thermal model
with Pb-Au experimental data would require, however,
weighted averages over all        events          with different
$dN^{ch}/dy$. For that one would still  need to define a model
describing the dependence of the effective initial volume on
the  impact parameter. In order to avoid  additional uncertainties
we compare here a thermal model with the CERES data corresponding   to
the highest multiplicity beams
          with $<dN_{ch}/dy> \sim 410$. For the most central
events, the transverse size of the initially created fire-cylinder
in Pb-Au is assumed to be of the order of the Au-radius.

For the isentropic longitudinal expansion the         temperature $T_0$,
the initial thermalization time $\tau_0$ and transverse size $R_A$ of
the initially created fire-cylinder in central A-A collisions
 can be related to the final state hadron
multiplicity  by~\cite{rudi}
\begin{equation}
{{dN}\over {dy}}\sim       \pi R_A^2 \tau_0 s(T_0)
\label{eq:murnf8}
\end{equation}
where $R_A \sim 1.2A^{1/3}$ is the nuclear radius and $s(T_0)$
the initial entropy density in the thermal medium.

>From Eq.1 one          establishes  and compares the initial
 thermal parameters for          S-Au and  Pb-Au collisions.
 CERES has measured $<dN^{ch}/dy>\sim 125$ and
  $<dN^{ch}/dy>\sim 410$ charged particles in  central collisions
  in S-Au and Pb-Au, respectively.
 Thus, assuming the same thermalization time for S-Au and Pb-Au
 one derives from Eq.1,
\begin{equation}
<{{dN^{ch}}\over {dy}}>^{S}/
<{{dN^{ch}}\over {dy}}>^{Pb}\simeq
({{A_S}\over {A_{Au}}})^{2/3} {{s(T_0^{S})}\over {s(T_0^{Pb})}}
\label{eq:murnf}
\end{equation}
This result indicates that the increase of the charged particle multiplicity
in central Pb-Au in comparison to S-Au collisions by factor $\sim$3.3 is almost entirely
described by the increase of the initial transverse size of the system
since $({\rm A_{Pb}}/{\rm A_S})^{2/3}\sim 3.35$. Thus, the  temperature  of the initially
created fire-cylinder in S-Au and  Pb-Au central collisions
is  the same within a few percent. Similar initial temperatures imply
similar
results for the dilepton yield normalized to the charged particle  multiplicity.
Thus, in terms of  a thermal model the dilepton data measured by CERES in
S-Au and Pb-Au for the most central collisions should coincide.
In fig.1 we see that the measured yield by CERES follows the above
expectations.
\begin{figure}
\center
 \psfig{figure=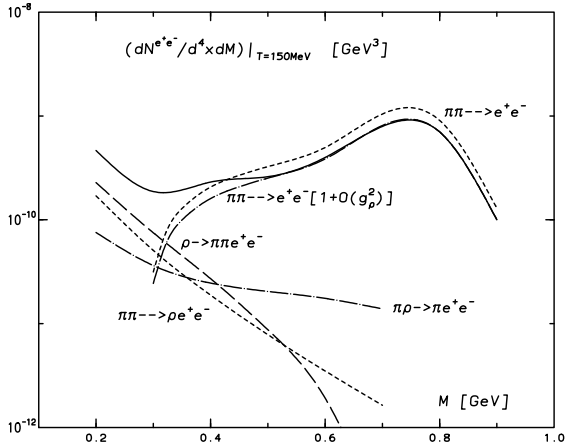,height=6.2cm}
\caption{Dilepton rate due to
 $\pi -\rho $ scattering as a function of {\it M} at fixed T=150 MeV.$^{14}$
The solid curve represents the sum of the real and the virtual process.
}
\label{fig:2   }
\end{figure}
The results of fig.1 show  that indeed thermal production
could be the source of the dilepton excess seen in the CERES experiment.
However, the Born term alone
can not explain  the characteristic
structure and the magnitude  of the excess seen in the
data. Similar conclusions
have been   obtained  in the previous studies~\cite{dinesh,ko,cassing,vesa,wambach,our,ko2,shur2}
 even when a more  complete model
for the expansion dynamics, reproducing not only the total pion     multiplicities but
also their measured p$_t$ and rapidity distributions has been applied.
This result is as well  independent on the assumption made about the nature of the initially created
thermal fireball, that is, whether it is   highly excited
hadronic matter or a quark-gluon plasma.
\begin{figure}
\center
 \psfig{figure=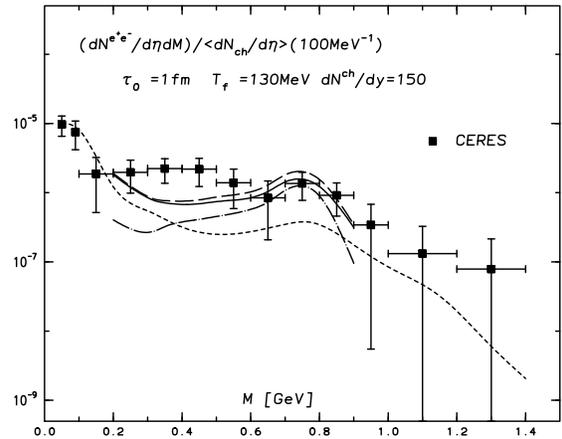,height=6.2cm}
\caption{Invariant-mass dielectron spectra measured by the CERES Collaboration
for the S-Au collisions in comparison with the thermal yield from $\pi -\rho$
interactions.$^{14}$ Dashed-dotted line describes the sum of all contributing
reactions from $\pi^\pm -\rho^0$ interactions. The dashed line,
calculated by CERES, represents the expected dielectron yield from
all known hadronic sources. The full-line is the sum of the contributions
described by the dashed-dotted and dashed curves. The long-dashed
curve includes the out of equilibrium effects.
}
\label{fig:3   }
\end{figure}
In fig.1 we have already seen the importance of the pion annihilation
process to partly understand the dilepton data.
 In a thermal medium, however, this basic reaction is not the only
 one  which  should be considered. For  example the contribution
                       from the two-body reaction
 $\pi^+\pi^-\to \rho\gamma^*\to \rho e^+e^-$ does not have a
  kinematical threshold at $2m_\pi$, and therefore  will             dominate
       pion annihilation
 for  $M\sim 2m_\pi$. This example indicates, that a more
 complete analysis of the low mass dilepton spectrum
 originating from     $\pi-\rho$  interactions is required.~\cite{our}

\section{Thermal dileptons     from $\pi -\rho $ scattering}
The thermal emission rate of heavy photons with invariant mass
$M$, energy $E$ and momentum $\vec q$ can be obtained from  the
photon self-energy tensor $\Pi_{\mu\nu}$ as follows~\cite{mclerran,weldon}
\begin{equation}
{{dR}\over {dM^2d^3q/E}} =
-{\alpha\over {24\pi^4 M^2}}n(E){\rm Im}\Pi^\mu_\mu (E, \vec q )
\label{eq:murnf6}
\end{equation}
where $n(E)$  is the Bose distribution function at temperature {\it T}.
The virtual photon self-energy is usually approximated
by carrying out a loop expansion to some finite order.
On the one-loop level one recovers the          expression
for the Born term $\pi^+\pi^-\to \gamma^* \to e^+e^-$,~\cite{mclerran}
which under the Boltzmann approximation and with $E^2={\vec q}^2+M^2$ reads:
\begin{equation}
{{dR}\over {dM^2{{d^3q}\over {E}}}}=
{{\alpha^2}\over {96\pi^4}} |F_\pi(M)|^2 \exp (-E/T)
\label{eq:murnf1}
\end{equation}
where $|F_\pi|$ is the pion form factor
\begin{equation}
|F_\pi(M)|^2 = {{m_\rho^4}\over {(M^2-{m_\rho^*}^2)^2
 +\Gamma_\rho^2m_\rho^2}}
\label{eq:murnf2}
\end{equation}
The parameters
$m_\rho=0.775$ GeV,
$m_\rho^*=0.761$ GeV and $\Gamma_\rho$= 0.118 GeV are       chosen
to get a reasonable description of the measured pion electromagnetic
form factor.~\cite{gale}

 To go beyond the one-loop approximation and to include the contribution
due to $\pi^\pm -\rho^0$ scattering  we adopt the  effective  Lagrangian~\cite{our,kapusta}
with the rho-meson   and electromagnetic
fields coupled to the pion current.
\begin{figure}
\center
 \psfig{figure=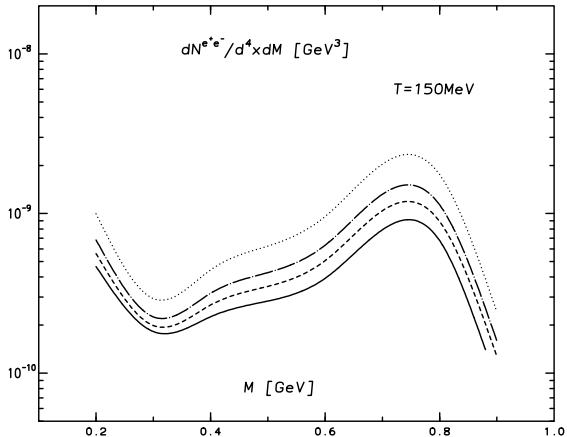,height=6.2cm}
\caption{Estimate of out of equilibrium effects to the dilepton rate up
to $O(g_\rho^2)$ order at fixed T=150 MeV:$^{14}$ solid curve corresponds to $\mu_\pi =0$,
dashed to $\mu_\pi =25,$
dashed-dotted to $\mu_\pi =50,$
and dotted
to $\mu_\pi =100$ MeV.}
\label{fig:4   }
\end{figure}
>From this  Lagrangian and using the closed-time-path formalism
the dilepton production rate has been calculated at the two-loop
level.~\cite{our} There are two types of            diagrams
contributing to the thermal dilepton rate from $\pi -\rho$ interactions. These are the diagrams
with real and virtual $\rho^0$ vector mesons.
The processes involving real  ${\rho^0}^,$s  are due to
$\pi\pi\to\rho\gamma^*$, $\pi\rho\to\pi\gamma^*$ and $\rho\to\pi\pi\gamma^*$
reactions.
The estimate of
 the resulting thermal rate    can be  found in Ref.[14].

  For heavy photon production the virtual contributions
  lead to $O(g_\rho^2)$ order corrections to the Born term.
  In the limit where $m_\rho >>T$ the  dilepton
  yield due to pion annihilation including  strong interaction $O(g_\rho^2)$ corrections
  can be estimated as,~\cite{our}
\begin{equation}
{{dN^{Born+virtual}}\over {dM^2d^4x}}\simeq
{{dN^{Born}}\over {dM^2d^4x}}[1-{7\over \pi}{{g_\rho^2}\over {4\pi}}
({T\over {m_\rho}})^2]
\label{eq:m1   }
\end{equation}
In fig.2 we summarize  the contributions to the thermal dilepton
rate originating from $\pi -\rho$ scattering.
It is clear that         dileptons with invariant masses
$M<0.45 $ GeV are mostly produced due to $\pi -\rho $
scattering
processes. It is also interesting to note that the $O(g_\rho^2)$
order corrections to the Born term are negative and relatively large.
It suggests that     resummations could be  required here.

The result in fig.2. shows that in a
 mesonic medium dileptons originating form
$\pi -\rho $ interactions have to be   necessarily included as they
lead to substantial modifications of the Born rate.

Applying Bjorken$^,$s model for the space-time evolution
one can  compare then the thermal dilepton production with the  experimentally measured  yield.
 In fig.3 we show the overall thermal dilepton rate from
 $\pi -\rho$  interactions including acceptance and kinematical cuts
 of the CERES experiment. In  fig.3 one can see
 that the thermal source for dielectron pairs can  only partly
 account for the excess measured by the CERES  Collaboration.
 In the model describing dilepton production due to $\pi -\rho$
scattering we have not included the possible coupling to the $A_1$
axial vector meson. However, the role and the  contribution of the  $A_1$ resonance
to thermal dilepton yield is extensively discussed in Ref.[17,26].

\subsection{Non-equilibrium effects}
In the previous section
discussing thermal dileptons we  assume
the free phase space thermal-distribution function for  all          particles.
In a high density
medium, however,  the particle properties
can be  substantially modified.~\cite{wambach,koch,gery1,gery,rob,song}
 This modification
could have a dynamical origin, or it could as well appear   due to non-equilibrium effects.
 In the following
we discuss  how the chemical off-equilibrium effects may influence
the soft dilepton yields.~\cite{our,zah,peter}
\par
Thermalization of a hadronic medium created in heavy ion collisions
should be    rather a fast process. Recent calculations in kinetic theory show
that
only few elastic particle collisions are  already sufficient   to maintain
thermal particle spectra.~\cite{gavin} The  chemical equilibration, in contrast,
could be much slower as it requires a detailed balance
between different reactions  with particle number changing processes.
The absence of chemical equilibrium in a pion medium can be effectively
taken into account by modifying the pion  distribution function~\cite{gavin}
\begin{equation}
n_{\vec \pi}(E       ,T)= (\lambda_\pi^{-1}  \exp {E/ T}-1)^{-1}
\label{eq:mm1   }
\end{equation}
with  $\lambda_\pi\equiv \exp (\mu_\pi/T)$ and $\mu_\pi$ being  the pion chemical potential.
Assuming   relative equilibrium between pions and rho-mesons
implies that $\mu_\rho =2\mu_\pi$.
If the pion fluid  were in chemical equilibrium, then naturally
$\mu_\pi$  would vanish.
\begin{figure}
\center
 \psfig{figure=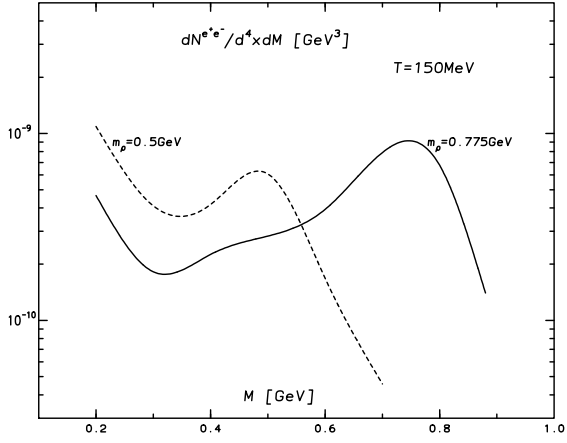,height=6.2cm}
\caption{
 Dilepton rate at T=150 MeV due to $\pi -\rho $  scattering
calculated with $m_\rho =0.775$ GeV (full-line) and
with $m_\rho =0.5  $ GeV  (dashed-line).
}
\label{fig:5   }
\end{figure}
The modification of  the dilepton production  rate due to $\mu_\pi\neq 0$
can be qualitatively verified taking as an example the Born term
for the pion annihilation process.
The dilepton production rate due to pion annihilation in a non-equilibrium
mesonic medium may be obtained in the
following form:~\cite{peter,jean}
\begin{equation}
{{dN} \over {d^4x dM^2({{d^3q}\over E})}}|_{\mu_\pi \neq 0}\simeq
{{dN} \over {d^4x dM^2({{d^3q}\over E})}}|_{\mu_\pi =0}
\times   F_h
\label{eq:m18   }
\end{equation}
where the equilibrium rate  is as in Eq.4 and the function  $F_h$
reads
\begin{equation}
\begin{array}{rcl}
&& F_h \sim
{T\over q}
{1\over {1-\exp [ -(E-2\mu_\pi )/T]}} e^{2\mu_\pi /T}  \\
& &  \times \ln {{{
 (e^{-\beta E_{+}}-e^{\beta \mu_\pi })
 (e^{-\beta E_{-}}-e^{\beta E}e^{\beta \mu_\pi })}
 \over
 {(e^{-\beta E_{-}}-e^{\beta \mu_\pi})
 (e^{-\beta E_{+}}-e^{\beta E}e^{\beta \mu_\pi })}}}      \\ & &
\end{array}
\label{eq:m5}
\end{equation}
where
$ E_{\pm}={1\over 2} [E\pm q(1-{{4m^2}\over {M^2}})^{1/2}]
$

The  $\mu_\pi$
is modifying    the   dilepton rate in Eq.8 in  two different
ways; first, due to the  factor  $\exp (2\mu_\pi /T)$ the overall
thermal production rate  is  enhanced independently on the dilepton
kinematics. Second, the appearance of the Bose-like
term in the denominator of Eq.9  enhances the  dilepton
production  when  $E\to 2\mu_\pi$.
This second feature is of particular interest as it leads to
an  excess of soft dileptons just above the $2m_\pi$ threshold.
\par
It is clear that not only pion annihilation but  all processes
involving  pions and rho mesons in the initial state are modified
in an over-saturated compared to an equilibrium  medium.
In particular the processes listed in the previous section,
arising from $\pi -\rho$ scattering
 are influenced by the  off-equilibrium effects.~\cite{our}
\begin{figure}
\center
 \psfig{figure=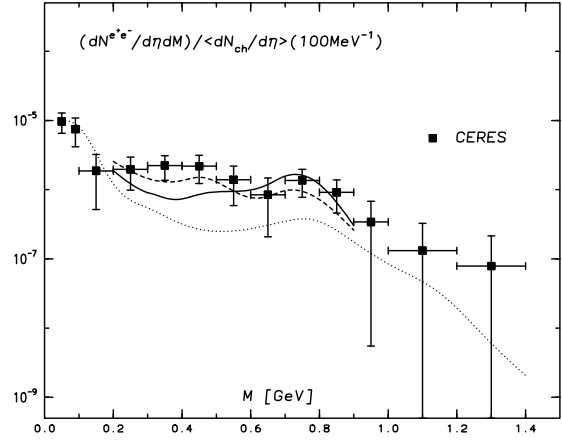,height=6.2cm}
\caption{
 Thermal dielectron yields in comparison
with     S-Au data. The full-line as in fig.3; the dashed-line
calculated  in a simple model for a dropping $\rho$-meson mass.
}
\label{fig:6}
\end{figure}
Discussing dilepton production in a thermal medium 
 requires     additional care with respect to   finite temperature modifications
of the rho meson mass and/or  decay width.~\cite{gery1,gery,rob}
The previous study of the rho selfenergy in a mesonic medium up to the one-loop order
has shown that in an equilibrium system the  medium effects on rho properties   are
rather modest.~\cite{gale}
 The situation can, however,  change in supersaturated pion~\cite{peter}
  or baryon rich~\cite{wambach,wolfgang}      matter.
In the first  case, for large values of $\mu_\pi$ and $T$ the increase of
the rho decay width may be significant  leading to the suppression of
dilepton production close  to the rho resonance peak.~\cite{peter} 

To include non-equilibrium effects in  the processes from
$\pi -\rho$ scattering, one would   just multiply {\ naively} the rate
    by a factor of $\lambda_\pi^2$ or
 $\lambda_\pi^3$ depending whether there is a $\pi\pi$ or $\pi\rho$  initial state.
 For sufficiently large $\mu_\pi$ the increase of the soft part of the dilepton
yield due to a non-equilibrium chemical potential is then
sufficient to explain the experimental data.
However, a more careful analysis shows~\cite{our} that it is not justified to simply multiply the equilibrium emission rate
for dileptons from $\pi -\rho$ scattering by   $\lambda_\pi^i$ factors.
This is mostly
due to the structure of the (one-loop) self-energy correction to the
pion propagator  which in the non-equilibrium medium leads to
an additional {\it pinch-singular}~\cite{our,tangi,le,last} term contributing to the overall dilepton rate.~\cite{our}
If the  deviation from equilibrium distributions is  small such that
$\delta\lambda \equiv \lambda_\pi -1 <1$, then the {\it pinch-singular} term
is  found as follows,~\cite{our}
\begin{equation}
{{dN^{pinch}}\over {dM^2d^4x}}\sim
-\delta\lambda  {{\alpha^2(g_\rho^2/4\pi )}
\over {24\pi^3}} \sqrt {{{\pi T^3}\over {2m_\rho^3}}}
{{m_\rho^3
e^{-{{{2m_\rho^2+M^2}\over {2m_\rho T}}}}}
\over {g_\rho^2/(4\pi e)T}}
\label{eq:aa   }
\end{equation}
Thus, since this term is  negative it  reduces the  naively expected
increase  of low mass dileptons due to a nonzero pion   chemical potential.

In fig.4 we plot the total rate, including the Born term with $O(g_\rho^2)$
corrections, for the non-equilibrium case characterized by different
values of the pion chemical potential $\mu_\pi$. We observe that
the rate is increasing with increasing $\mu_\pi$. However,
naively one would expect an increase by a factor of $\lambda_\pi^2$,
whereas the results in fig.4 show much lower enhancements.
Changing  $\mu_\pi $ from $\mu_\pi =0$ to $\mu_\pi =100$ MeV only an effective
increase by a factor 2 results in fig.4, contrary to a factor
4 expected from  $\lambda_\pi^2$. This is mostly because of the negative contribution of
the $pinch-singular$ term in Eq.10. From this we stress the importance of taking
into account the non-trivial term (Eq.10), which is   traced back to
the structure of the pion propagator.

In fig.3 we show the thermal rate (long-dashed line)
calculated in the longitudinally expanding fire-cylinder
assuming deviations from chemical equilibrium
with a   value of the chemical potential
of $\mu_\pi =100$ MeV. Indeed, we observe an increase of the rate
below the rho peak. This increase, however, is rather modest
due to the $pinch-singular$ term.
In addition, the off-equilibrium distributions of pions and rho mesons
imply lower initial temperature and a shorter lifetime
of a thermal system. All these effects are the reason of the small
increase of the low mass yield when going from an equilibrium
to a non-equilibrium medium.

\subsection{Dropping vector-meson masses}
The modification of the pion and rho meson properties by a
finite     chemical potential although implies some increase of the low mass
dilepton rate but still it is not sufficient to explain the data.
 In general, all the models discussed in  literature,
 until now,~\cite{dinesh,ko,cassing,vesa,wambach,our,ko2,shur2}
lead to the similar conclusions   that  the  conventional
sources for dilepton production
 can only partly account for
the excess measured by the CERES Collaboration.
One possible explanation of why the conventional approach fails
may be due to
the fact that hadronic properties in a high density  medium may be
modified and these effects have to be taken into account. Particularly relevant here would be  the modification
of the rho meson mass.~\cite{ko,gery1,gery,rob} The properties of the rho meson
in a high density medium are, however, theoretically
not      under complete control. The results of lattice gauge theory calculations in quenched QCD~\cite{lgt} as well
as the calculations based on the effective chiral Lagrangian~\cite{song2}
show  that the rho meson mass at finite temperature remains almost
unchanged. On   the other hand the results of gauged linear sigma
models suggest an increase of the rho meson mass with temperature.~\cite{rob}
Finally, Brown and Rho~\cite{gery1} and Adami and Brown~\cite{gery1} have shown that
the restoration of chiral symmetry implies a decreasing rho mass
in dense matter.
In the following we assume a decreasing rho meson mass in a medium
and discuss how this       influences the dilepton yield.

In fig.5 we compare the dilepton rate due to $\pi -\rho$ scattering for two different
values of $m_\rho =0.77$ and $m_\rho =0.5$ GeV keeping the rho decay width unchanged. As expected decreasing $m_\rho$ leads to the shift
in the position of the resonance peak, increasing in this way the production
of low mass dileptons. This is the feature
which is required to understand the low mass dilepton excess.
It is also interesting to note that the height of the peak at $M=m_\rho$
is a decreasing function of mass. This result is entirely
due to the $O(g_\rho^2)$ term          which as seen in Eq.6 to be  strongly
dependent on the value of $m_\rho$. This also means that
assuming a decreasing vector meson mass we have to worry about
the strong interaction corrections  to the purely electromagnetic
pion annihilation  process, e.g. when using the Lagrangian from Ref$^,$s.[14,25].
\par
In order to develop a shift in the position of the rho meson peak
in the dilepton yield, the  rho meson  should stay off-shell a
substantial amount of time
during the evolution of the hadronic medium. This can be the case if
 $m_\rho$ is a slowly varying function of temperature
and/or baryon density~\cite{ko,cassing} or if the system stays sufficiently
long close to the chiral  symmetry restoration phase transition point.~\cite{shur2}
To illustrate how a dropping rho meson mass modifies the yield
we adopt  a  very
 simple parameterization
of the  T-dependence of $m_\rho$. We assume that up to  half of the lifetime
of the thermal medium the rho meson stays with reduced mass $m_\rho^*\sim 0.5$ GeV
and then very quickly recovers its on-shell value $m_\rho\sim 0.77$ GeV.
>From this simple example, one can already see in fig.6  that a  dropping
rho meson mass may be the source of the structure of the rate seen
in the CERN  experiments.
\par
The quantitative explanation of the data with a dropping vector meson
mass has been shown first   by Li, Ko and Brown.~\cite{ko}  In fig.7 we show the
results of Ref.[10] obtained in terms of a hadronic transport model for expansion
dynamics and the parameterization of the vector meson masses    deduced  from
    Brown-Rho scaling.~\cite{gery1} The same model shows also   very good agreement
with HELIOS/3 dimuon data.~\cite{ko2} A similarly good agreement
with the CERN dilepton data is  obtained in different dynamical models
for dilepton production in A-A collisions when assuming  dropping vector meson masses.~\cite{cassing}
Until now including  the decrease of  vector meson masses  seems to be the best
way to understand the properties, shape and excess of low mass dilepton
yields reported  by CERN experiments.
\par
\begin{figure}
\center
 \psfig{figure=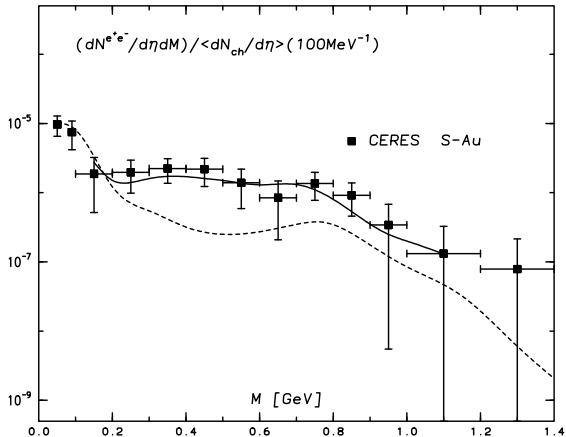,height=6.2cm}
\caption{Thermal dilepton yield with a dropping $\rho$-meson mass from
Ref.[10] in comparison with the CERES S-Au data.
}
\label{fig:8   }
\end{figure}
\section{Outlook and Conclusions}
We have discussed thermal dilepton production in a hot mesonic
medium in the context of recent CERN experimental data by the  CERES
Collaboration for S-Au and the most central Pb-Au collisions at SPS energies.
The calculation includes all production processes due to
$\pi -\rho$ scattering originating from two-loop
approximations of the virtual photon self-energy.
We argue that in an equilibrium model $\pi -\rho$
scattering is an important source of low mass dileptons. Adding  it
 together with the Born term for pion annihilation one can
partly explain the dilepton excess measured in the CERN  experiment.
Next, the off equilibrium contributions to the dilepton
rate due to a non zero pion chemical potential have been found
to imply only a modest modification of the dilepton production.
This is mostly due to the
structure  of the non-equilibrium pion propagator.
Finally, the modification of dilepton yields due to in medium effects
on vector meson properties is discussed. We emphasize that a dropping rho
meson mass is  the best way, until now,  to provide a quantitative
explanation of the recently observed low mass dielectron enhancement
in S-Au and Pb-Au collisions.
The concept of a dropping vector meson mass should be, however,
cross checked with the upper limit of direct photon yields
measured in S-Au collisions by the WA80 Collaboration.~\cite{santo} It is not
excluded that the thermal photon yields calculated with a dropping rho meson mass
may overwhelm the experimental upper limit.
Additional attention  should be given  to the higher  $O(g_\rho^2)$  corrections
to  the   $\pi^+\pi^-\to e^+e^-$  annihilation process.
We estimate  that  the $O(g_\rho^2)$ term is negative and thus reduces the Born term.
This reduction of the dilepton yield is particularly important in case of a
dropping  rho resonance mass.

\section*{Acknowledgments}
One of us (K.R.), acknowledges stimulating discussions with the members
of the CERES Collaboration and support of the Gesellschaft f\"ur Schwerionenforschung (GSI Darmstadt)
and of the KBN 2-P03B-09908. (K.R.) also acknowledges discussions
with B.~L. Friman, J. Rafelski,  H. Satz and D. K. Srivastava.
M.D. is supported by DFG.

\end{document}